\documentstyle[12pt, psfig]{article} 
 
\newcommand{\be}{\begin{equation}} 
\newcommand{\ee}{\end{equation}} 
\newcommand{\bea}{\begin{eqnarray}} 
\newcommand{\eea}{\end{eqnarray}}

\begin{document} 
\begin{center} 
{\LARGE \bf Oscillating shells: A model for a variable cosmic object}\\ 
\vskip1cm 
{\bf Dar{\'\i}o N{\'u}{\~n}ez}\\ 
\vskip5mm 
Instituto de Ciencias Nucleares, UNAM, \\ Circuito Exterior CU, A.P. 70-543, 
M{\'e}xico, D. F. 04510,  M{\'e}xico.\\  
e-mail nunez@nuclecu.unam.mx\\ 
\end{center} 
 
\begin{abstract} 
{\it A model for a possible variable cosmic object is presented. The model 
consists of a massive shell surrounding a compact object. The gravitational 
and self-gravitational forces tend to collapse the shell, but the internal 
tangential stresses oppose the collapse. The combined action of the two 
types of forces is studied and several cases are presented. In particular, 
we investigate the spherically symmetric case in which the shell
oscillates radially around a central compact object.}
\end{abstract} 
 
\vspace{7mm} 
Subject headings: Thin Shell, Oscillation, Dynamics-Stars

\vspace{7mm} 
{PACs number: 97.60L} 
\vfill 
\eject

\section{Introduction} 
\vspace{7mm} 
Since the pioneering works of Ostriker and Gunn (Ostriker and Gunn, 1977), 
the 
study of the motion of thin clouds surrounding an exploding star has been the 
basis for understanding the dynamic process occurring after a supernova 
explosion, the motion of a blast wave (Sato, 1993) and several other related 
phenomena. These studies were done in Newtonian theory, where 
an equation of motion derived from Newton's second law was used, introducing 
the forces that act on the shell. For the case of a supernova we have as the 
interacting forces the gravitational force, the force due to the radiation 
pressure emitted by the star, and a friction type force due to the 
interaction of the shell with the interstellar medium. The deduction is 
clear and 
straight-forward and the picture obtained is quite useful and has 
a good degree of accuracy for several cases in describing the main processes 
occurring during the motion.  
 
On the other hand, in General Relativity, the relativistic 
version of the equation of motion of a shell has its origins in studies of 
the conditions needed to join two regions of space-time, the so called 
matching conditions. Initially, the Lichnerowicz matching conditions 
(Lichnerowicz, 1955) gave conditions on the boundary between two regions 
of the 
space time, where the boundary was just the end of one region and the 
beginning of 
the next. Israel (Israel, 1966, 1967) gave a different interpretation to 
this boundary, allowing it to be a thin shell of matter, thus the two regions 
of the space time are matched on this boundary, the thin shell, obtaining new 
matching conditions, for which, when the shell parameters tend to zero, the 
Lichnerowicz ones are recovered. Israel's matching conditions for two spaces 
can be interpreted as the equations that govern the motion of the shell in 
order to consistently separate the two given space times, that is, can be 
read as the 
equations of motion for the shell, which has had very different 
applications, such as 
allowing the study of collisions of shells (N{\'u}{\~n}ez et al. 1993). 
Recently  
N{\'u}{\~n}ez and Oliveira ( N{\'u}{\~n}ez and Oliveira 1996) explicitly 
obtained the general 
relativistic version of the Newtonian equation of motion. Thus, both theories 
are now on the same footing for analyzing the motion of a shell.  
 
Even though  the Relativistic analysis demands for its formulation a more
complicated  
mathematical machinery and often the results are negligible corrections to 
the Newtonian description, the theory has several conceptual advantages and 
applications that make it worth considering as an alternative approach to 
analyzing the motion of shells. It is definitely necessary to use it in cases 
where the gravitational fields are strong, with high values of the density 
compared to the pressure or when the velocities are close to the velocity of 
light. With respect to the motion of the shell, it is  necessary to use the 
Relativistic formalism when the shell surrounds a black 
hole or when the shell itself is very dense and moves at high speeds. The 
model presented in this work belongs to this last case. Moreover, using the 
fact that in General Relativisty the explicit equations relating 
the geometrical quantities (the ``forces" of the Newtonian theory) with the 
matter energy distribution in the space are given (the Einstein equations!), 
the set of equations of motion can be reduced to a set of first order 
differential equations, that is, a first integral of the equations of motion
can be obtained, which in the Newtonian limit is related to the energy of the
system. The direct deduction of this first integral for the case of spherical 
symmetric space times with a matter shell of perfect fluid was done by 
N{\'u}{\~n}ez and Oliveira, (N{\'u}{\~n}ez and Oliveira 1996).  
 
In the present work we use this set of first order differential equations plus
an equation of state for the surface density and the tangential pressure 
of 
the shell, in order to formulate a model in which the shell moves due to the 
combined action of the tangential stress and the gravitational attraction. 
With an appropriate selection of the parameters, different types of motion 
can be described. We present four cases of motion, an indefinite expansion of 
the shell starting from rest, which can be seen as an explosion; two cases 
where the shell oscillates and a case where the shell is static. The 
oscillation case is particularly interesting because it implies a change in 
the shell's surface density which in turn could be related to a change 
in temperature, which would look like an object of variable magnitude to a 
distant observer. In all cases, the values of the parameters of the shell 
obtained, the tangential pressure and the surface energy are large. They 
are the two dimensional analogy of the values of a neutron star, so this 
models represent the motion of a sort of ``neutron shell". Finally, we have to 
call the reader's attention to the fact that the equation of state used 
in the derivation
of the oscillatory motions and the expansion one is quite unusual and 
it is unlikely that it describes a realistic type of matter, so the model
should be taken as a first approach. This is not
so for the static case, where the result is valid for all equations of state.
 
Even though in order to analyze the oscillation of an object such as a Cepheid
star one would need to take into account more parameters, we think that the 
type of analysis presented here is in the correct direction in obtaining an 
analytic model for such objects. 
 
The paper is organized as follows: In the next section we present the 
equations of motion. A detailed discussion of the projections of the 
geometrical 
quantities form both spaces on the shell and the deduction of the junction 
conditions is presented in the appendix. We also present a brief analysis
of the stability of the shells under radial perturbations. In section 3 we 
introduce an 
expression for the tangential pressure on the shell, 
obtain the equation of state
and construct the model; presenting and  analyzing several cases for different
sets of parameters. We also include the static case which, as we said, is 
independent of the equation of state. Finally in section 4 we give some 
conclusions and mention 
possible directions for further research.

\section{Equations of Motion and Stability Analysis} 
 
As was mentioned above, in the General Relativistic formalism the motion 
equations of the shell are obtained as matching conditions between the two 
space-times that the shell separates. In the case when these space-times are 
spherical symmetric and the matter - energy of the shell is described by 
a perfect fluid, the equations of motion reduce to a set of two coupled 
first order differential equations. 
 
The general form of the line element of a spherical symmetric space time is 
given by: 
\be 
ds^2=-e^{\psi}dv((1-{{2\,m}\over r})\,e^{\psi}dv - 2dr)+r^2(d\theta^2+
\sin^2\theta\,d\phi^2), 
\label{eq:lel}\ee 
where $v$ is the advanced null coordinate, $m$ and $\psi$ are arbitrary 
functions of $v$ and r. We are using the usual units in General Relativity, 
where the speed of light and the gravitational constant are equal to unity, 
but in the concrete examples we will give the quantities in the international
system of units as well. For this symmetry the equations of motion are, the 
mass-energy conservation equation: 
\be 
\dot{M}+p\,\dot{A}=A\,[T_{\mu\nu}n^\mu\,u^\nu],\label{eq:mot1} 
\ee 
where $M$ is the mass of the shell (not necessarily constant!), $p$ is its
two - dimensional tangential pressure (it is {\underline not} 
the ordinary 3 - 
dimensional pressure) and $A=4\pi\,R^2(\tau)$ is the area of 
the shell, related 
to the shell's density $\sigma$ by $\sigma={M\over A}$, (see Appendix). A 
quantity in squared brackets stands for the difference of that quantity 
evaluated outside the shell, $+$, minus the quantity evaluated inside, $-$.  
Dot, `` $\dot{} $'' denotes differentiation with respect to the proper time of
the shell, $\tau$, given in Eq.~(A.8).
 
The second equation is the dynamic equation for the radius of the shell: 
\be 
\dot{R^2}+V(R)=0, \label{eq:mot2} 
\ee 
where $V(R)$ is the potential given by 
\be 
V(R)=1-({{m_+-m_-}\over{M}})^2-{{m_++m_-}\over{R}}-({{M}\over{2\,R}})^2,
\label{eq:pot} 
\ee 
 
The Einstein equation for the line element given above are for each space 
time 
\bea 
m_{,v}= 4\,\pi\,r^2\,{T^r}_v,& & m_{,r}= -4\,\pi\,r^2\,{T^v}_v, \nonumber\\ 
& \psi_{,r}= 4\,\pi\,r\,T_{rr},& 
\eea 
 
To completely determine the motion of the shell, one has to supply an
equation of state for the 
matter of the shell which relates the tangential pressure 
with the superficial density, $p=p(\sigma)$. In this way, the dynamic problem 
for determining the motion of the shell between two given spaces consist of 
two first order coupled differential equations. 
 
A typical problem could be posed as follows: First determine
where the shell is 
going to be moving, that is, give the metric tensors $g_{\mu\nu}^+$, and 
$g_{\mu\nu}^-$ of the two regions of the space-time, ${\cal M}_+$ and 
${\cal M}_-$ respectively, which is equivalent to giving the matter - energy 
distribution of each region of the space time through the Einstein's 
equations . Second we choose a equation of state for the matter on the shell, 
we specify 
what it is composed of, $p=p(\sigma)$.  This information completely 
specifies all the parameters in the equations of motion, so we can proceed to 
their analysis. 

Notice that the formalism to study the motion of the shell 
has been developed to the point
that no more deduction of equations is needed, we need only define the 
parameters and proceed directly to study the equations of motion.

Finally, it is of interest to say some words about the stability of
the shells. We will restrict ourselves 
to the stability analysis with respect to radial 
perturbations, which are most important because of the spherical symmetry 
and describe how the shell reacts to the influence of radial deviations.

We start from the dynamic equation for the radius of the shell, 
Eq.~(\ref{eq:mot2}), and perform a variation $R(\tau)\to R(\tau) +
\delta R(\tau)$, which, to first order in $\delta R(\tau)$ implies:
\be
\dot{R}^2(\tau) \to \dot{R}^2(\tau) + 2\,\dot{R}(\tau)\,(\delta R(\tau))^., 
\nonumber
\ee
and for the potential, $V(R)$, we have that $V(R) \to
V(R +\delta R)$, and a Taylor expansion around $V(R)$ to first
order in $\delta R$ gives (in what follows we do not show explicitly the
dependence on $\tau$):
\be
V(R) \to V(R) + {{\partial V}\over{\partial R}}|_R \, \delta R. \nonumber
\ee

Substituting these last two equations in Eq.~(\ref{eq:mot2}) and using the
fact that we are perturbing a solution of this equation, we
get:
\be
2\,\dot{R}\,(\delta R)^. + {{\partial V}\over{\partial R}}|_R \, \delta R =0, 
\label{eq:1per}\ee
which is the equation for the evolution of the radial perturbation. Again 
using Eq.~(\ref{eq:mot2}), expressed as $\dot{R} =\sqrt{-V}$, 
which on integration yields:
\be
\delta R = {\rm exp}\,\int {{\partial \sqrt{-V}}\over{\partial R}}|_R\,d\,
\tau. \label{eq:pert}
\ee

Now the integrand, using Eq.~(\ref{eq:pot}), has the form:
\be
{{\partial \sqrt{-V}}\over{\partial R}}|_R=
{{-{{2\,(m_++m_-)\,R+M^2}\over{2\,R^3}} + 
({{M^4-2\,(m_+-m_-)^2\,R^2}\over{2\,M^3\,R^2}})
{{\partial M}\over{\partial R}}
\over{2\sqrt{-V}}}} .
\label{eq:dpot}\ee

In order to proceed further we would need to specify the model chosen and then
obtain explicitly R and M as functions of $\tau$, perform 
the integration and study how the perturbation behaves. Fortunately, some 
general remarks can be made about the integrand so something can be said 
even without an explicit solution. We will return to this point in the
next section.

\vspace{7mm} 
\section{The Model} 
 
In the present work we want to show the action of two combined forces acting 
on the shell and how their interaction produces several types of motion, 
including oscillatory motion. We will construct a simple model where there
will be vacuum outside the shell, and a compact object in vacuum inside of it,
that is, we will take the outside and inside space times to be described by 
the Schwarzschild metric, with  constant gravitational masses $m_+$ and $m_-$
respectively, and  $T_{\mu\nu}^\pm =0$. This will generate a gravitation 
attraction
towards the center, so it accounts for the inward force.  
 
With respect to the outward force, consider that, as it collapses the 
density of the shell grows, which in turn produces an increase in the 
tangential pressure, which generates a roman arch type of 
force that opposes the 
collapse. In the present model, we will represent the growth of the tangential
pressure between particles of the shell with dependence on the radius of the 
shell by the expression 
\be 
p=p_0\,e^{-\kappa\,R},\label{eq:pr} 
\ee 
with $p_0$ and $\kappa$ constant parameters, and $R=R(\tau)$ is the radius of 
the shell at the time $\tau$, the proper time of the shell. 
 
With these suppositions the mass-energy conservation equation of the shell, 
Eq.~(\ref{eq:mot1}), becomes 
\be 
\dot M = -8\,\pi\, p_0\,R\,e^{-\kappa\,R}\dot R, 
\ee 
which can be solved in terms of $R$: 
\be 
\tilde M=\tilde M_A + {{8\,\pi\,p_0}\over{\kappa^2\,M_\odot}}\,(1
+\kappa\,R_\odot\,\tilde R)\,e^{-\kappa\,R_\odot\,\tilde R}, \label{eq:mas1} 
\ee 
with $\tilde M_A=const.,$ the ``dust mass" of the shell, and we are expressing 
the masses as multiples of the Solar mass, $M_\odot$, and the radius $R$ 
as multiples of the solar radius $R_\odot$, that is $M=M_\odot\,\tilde M$,
$R=R_\odot\,\tilde R$. $\tilde M$ and $\tilde R$ are unitless. We can go 
further and obtain an equation of state for the 
matter of the shell,  
\be 
\sigma = {{M_\odot\,\tilde M_A\,\kappa^2 + 8\,\pi\,(1
-\ln({p\over{p_0}}))\,p}\over{4\,\pi\,({\rm ln}({p\over{p_0}}))^2}}. 
\label{eq:den} 
\ee 
 
This equation of state is rather unusual, but it is well behaved 
for 
$p>p_0$, which is always the case in the motion studied, since
the shell never 
reaches a zero radius. Besides, as will be shown below in the examples, 
it does not violates the energy conditions, namely 
$|p|< \sigma$ is always satisfied 
so the matter is not of an ``exotic" type. Still, we have to agree
that it is unlikely that the matter defined by such a equation of state
would be realistic, so the model presented here should be taken just 
as a first approach.
 
In the dynamic equation for the shell, Eq.~(\ref{eq:mot2}), the potential, 
Eq.~(\ref{eq:pot}), has the gravitational masses  $m_-$ inside and
$m_+$  outside, which are constants, and the gravitational 
mass of the shell is given 
by Eq.~(\ref{eq:mas1}). So in terms of the solar parameters we have 
for the potential the following expression: 
\be 
V= 1-({{\tilde m_+-\tilde m_-}\over{\tilde M}})^2-({{M_\odot}\over{R_\odot}})\,
{{\tilde m_++\tilde m_-}\over{\tilde R}}-({{M_\odot}\over{R_\odot}})^2\,
({{\tilde M}\over{2\,\tilde R}})^2,\label{eq:pot1} 
\ee 
and with the units we are using, $G=c=1$, both the Solar mass and the Solar 
radius are in 
length units, $M_\odot=1.473\,*\,10^5 \,{\rm cm}$, and 
$R_\odot=6.95\,*\,10^{10}\,{\rm cm}$. The 
constants $\kappa$ and $p_0$ have units of ${\rm cm}^{-1}$. 
 
The dynamic equation,  Eq.~(\ref{eq:mot2}), is the expression for the total 
conserved energy of the shell which, as it is characteristic in General 
Relativity, does not have 
an arbitrary value but a fixed one, (in our case zero), so the 
kinetic energy is equal to minus the potential energy for all the motion. 
From this fact we can conclude several properties of the potential term. 
First, as the kinetic energy has always to be positive, the potential in a 
well defined 
motion has to be negative over all the range of radius. Also, wherever the 
shell stops, the potential energy term has to be equal to zero at those points
and vice versa. A minimum for the potential corresponds to a maximum of the 
kinetic energy and of course there can be only one minimum between the 
turning points and the second derivative or the potential with respect to the 
radius has to be positive.  
From inspection of the potential equation we can also conclude that far from 
the shell, the leading behavior of the potential is 
$1-({{\tilde m_+-\tilde m_-}\over{\tilde M_A}})^2$. Finally, from the relation
between the Solar mass and the Solar radius we expect a relation between the 
values of the radius, $R$, and the values of the masses, $m_+, m_-, M$ of the 
order of $10^6$. Thus the problem is posed as a five-parameter dynamic 
problem. 
 
With these general considerations, even though the motion equation can not be 
solved analytically, we can search for parameter sets which determine 
what motions could be described by our model.  
 
As a matter of fact, there is a large range of values of the parameters for 
which several types of motion can occur. We present three cases, two of 
oscillatory motion, one with masses of the order of unity, and another with 
radius of the order of unity, and a case with indefinite expansion 

Finally, we present a static case, in which the shell is at rest.
This case is particularly interesting because the mass energy conservation 
equation is satisfied directly and we do not need to specify an equation
of state and the results for this model hold for all equations of state.

With respect to the stability analysis, we can give also some general
remarks. As mentioned above, the motion is defined for a range of radius
where the potential is negative, so the integrand for the radial perturbation,
Eq.~(\ref{eq:pert}), is real for the range where the motion is taking place
so the perturbations either increase exponentially or decrease exponentially, 
but we do not expect oscillatory type of perturbations. For the
cases where the shell oscillates or expands, we would need the explicit
solution of the motion to say something more definite. Again, this is not the 
case for the static model where the perturbation analysis has to be
taken to the second order of the perturbations, where the perturbation
equation can be solved as we show below.

Now we proceed to present the different cases:
 
i) With values for the masses of the order of solar masses, we take for the 
Schwarzschild mass outside, $\tilde m_+ = 1.1$, that is 
$m_+ = 2.18\,*\,10^{33} {\rm g}.$; 
for the Schwarzschild mass inside, $\tilde m_- = 0.5$, that is 
$m_- = 9.93\,*\,10^{32} {\rm g}.$; 
for the gravitational ``dust mass" of the shell, 
$\tilde M_A = 0.603$, that is $M_A = 1.198\,*\,10^{33} {\rm g}.$; 
for the pressure constant 
$p_0=10^{-3} cm^{-1}$; and for the constant 
$\kappa=2.8\,*\,10^{-6} {\rm cm}^{-1}$.  
 
These values generate a potential of the form shown in Figure 1. Notice 
the region below the R-axis, where the potential takes negative values which, 
as we explained above, is the region where the motion is allowed. We have two 
crossing points with the R-axis where the potential equals zero, so the 
motion is bounded by these two values of the radius, which are 
$\tilde R_{\rm min}=9.0847\,*\,10^{-5}, 
(R_{\rm min}=6.313\,*\,10^{6} {\rm cm}.)$; 
and 
$\tilde R_{\rm max}=3.415\,*\,10^{-4}, 
(R_{\rm max}=2.373\,*\,10^{7} {\rm cm}.)$. 
At these 
extremes the shell stops and starts moving in the oppsosite direction; 
in Figure 2 we present 
a graph of the shell's velocity, reminding the reader that these velocities 
are given as factor of the velocity of light which in these units has a value 
of 1, ( $c=3\,*\,10^{10}{{\rm cm}\over{\rm sec}}$). Notice that the 
maximum velocity of 
the shell occurs closer to the minimum radius, where the tangential pressure 
acts strongly and decelerates the shell, 
finally stopping it. At that point the
velocity reaches a maximum with a value of the order of 0.14. Taking the 
average velocity to be roughly half this maximum, $\overline v =0.07$, and 
recalling that the one-way distance covered is 
$d=R_{\rm max}-R_{\rm min}=17.4\,*\,10^6 {\rm cm}.$, we obtain 
for the period of the 
oscillation, $T=1.65\,*\,10^{-2} {\rm s}.$  
 
Finally, from the expression for the surface density, Eq.~(\ref{eq:den}), 
and from that for the tangential pressure, Eq.~(\ref{eq:pr}), we can obtain 
the range of values over which these functions vary: 
\bea 
\sigma & \in  & [1.798\,*\,10^{-10}, 1.254\,*\,10^{-11}]\,{\rm cm}^{-1} 
= \nonumber \\&& 
[2.426\,*\,10^{18}, 1.693\,*\,10^{17}]\, {{\rm g}\over{\rm cm^2}}, \\ 
p &  \in & [2.0998\,*\,10^{-11}, 1.3758\,*\,10^{-32}]\,{\rm cm}^{-1} 
=  \nonumber \\ && 
[2.549\,*\,10^{38}, 1.670\,*\,10^{17}]\, {{\rm dyne}\over{\rm cm}}, 
\eea 
where in the brackets we give the values that the function takes at the 
minimum radius and at the maximum. We want to stress the fact that in the 
whole
range the strong energy condition, $\sigma >|p|$, holds which means that the 
energy density is positive definite for all observers so, as we 
mentioned, even though the equation of state is unusual, 
it does describe a 
well defined type of matter. Finally, we remind the reader that these 
values of density and pressure are defined on a surface, so the comparison of 
their magnitudes with quantities defined on volumes is not well posed. 
Nevertheless, we can say that we are talking about large densities and 
stresses, the ``neutron shell" that we mentioned. 
\begin{figure*}
\hspace{5cm}
\psfig{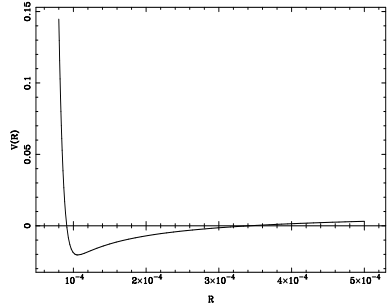}
\caption[]{\label{f1}
Potential for the motion of the shell in the case of masses of the 
order of solar masses. $\tilde R$ is given a multiples of the solar radius, 
the potential $V(\tilde R)$ has no units.}
\end{figure*}

\begin{figure*}
\hspace{5cm}
\psfig{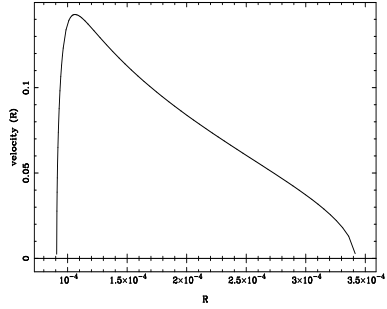}
\caption[]{\label{f2}
Shell velocity for the potential shown in Figure 1. The velocity 
is unitless, as multiples of the speed of light, $c=1$.}
\end{figure*}


ii) As was discussed above, for large radius the potential tends to a constant
value given by $1-({{\tilde m_+-\tilde m_-}\over{\tilde M_A}})^2$, so choosing
for the outside Schwarzschild mass, $\tilde m_+ = 1.5$, (
$m_+ = 2.98\,*\,10^{33} {\rm g}.$), for the inside Schwarzschild mass, 
$\tilde m_- = 0.73$, ($m_- = 1.45\,*\,10^{33} {\rm g}.$); 
and for the gravitational 
``dust mass" of the shell, $\tilde M_A = 0.74$, ($M_A = 1.47\,*\,10^{33} 
{\rm g}.$), 
the potential tends to the value $V\to -4.05\,*\,10^{-2}$; that is, it expands 
indefinitely with a constant velocity of 
$v\to 0.201 c =6\,*\,10^4{{\rm Km}\over{\rm s}}$. This behavior  
represents an explosion.
With the values of $p_0=5\,*\,10^{-4}\, {\rm cm}^{-1},\, 
\kappa=2.38\,*\,10^{-6}\,{\rm cm}^{-1}$, the 
potential is shown in Figure 3. The present case can be interpreted as a shell
starting from rest at a radius of $R_o=9.34\,*\,10^{-5}=6.473\,*\,10^6 
{\rm cm}$ 
from the 
center, and the tangential pressure expels it in such a way that the 
gravitational attraction cannot stop the motion, so the shell continues 
expanding indefinitely, tending towards a uniform motion with constant 
velocity. 
\begin{figure*}
\hspace{5cm}
\psfig{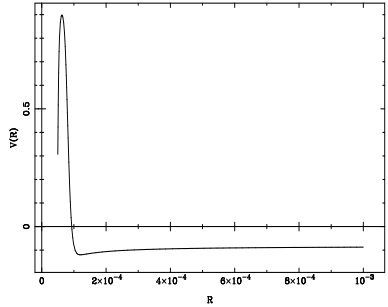}
\caption[]{\label{f3}
Potential for the case of an indefinite motion, the shell starts 
from rest at a radius of $\tilde R_o=9.314*10^{-5}R_\odot$.}
\end{figure*}

\vfill
\eject
 
iii) Choosing the parameters in such a way that the shell oscillates within 
values of the  radius of the order of a solar radius, we take 
$\tilde m_+ = 1.3\,*\,10^5, \tilde m_-=4.45\,*\,10^4, \tilde M_A=10^5 
(m_+ =2.59\,*\,10^{38}\,{\rm g}., m_- =8.51\,*\,10^{37}\,{\rm g}. 
M_A =1.989\,*\,10^{38}\,{\rm g}. )$, 
and taking $p_0=5\,*\,10^{14}\,{\rm cm}^{-1}, \kappa=7.41\,*\,10^{-10}\,
{\rm cm}^{-1}$. In this way we 
obtain a potential which again describes oscillatory motion. The region of 
the potential with negative values is presented in Figure 4. The minimum and 
maximum radius of the oscillation, the turning points, are 
$R_{\rm min}=1.2321\, R_\odot=8.56\,*\,10^{10}\,{\rm cm}., \, 
R_{\rm max}=1.4045 \, R_\odot = 9.76\,*\,10^{10}\,{\rm cm}$ which, as 
we wanted, are of the order of a Solar radius. 

A graph of the shell's velocity is presented in Figure 5. Notice that the 
maximum velocity of the shell is $v_{\rm max}=0.15 c$ 
(which is of the order of 
the maximum velocity for the first case presented). Taking the average as 
$\overline v = 0.07 c$, we obtain that the oscillation period for 
this case is $T=31.5 {\rm s}$. 
\begin{figure*}
\hspace{5cm}
\psfig{figure=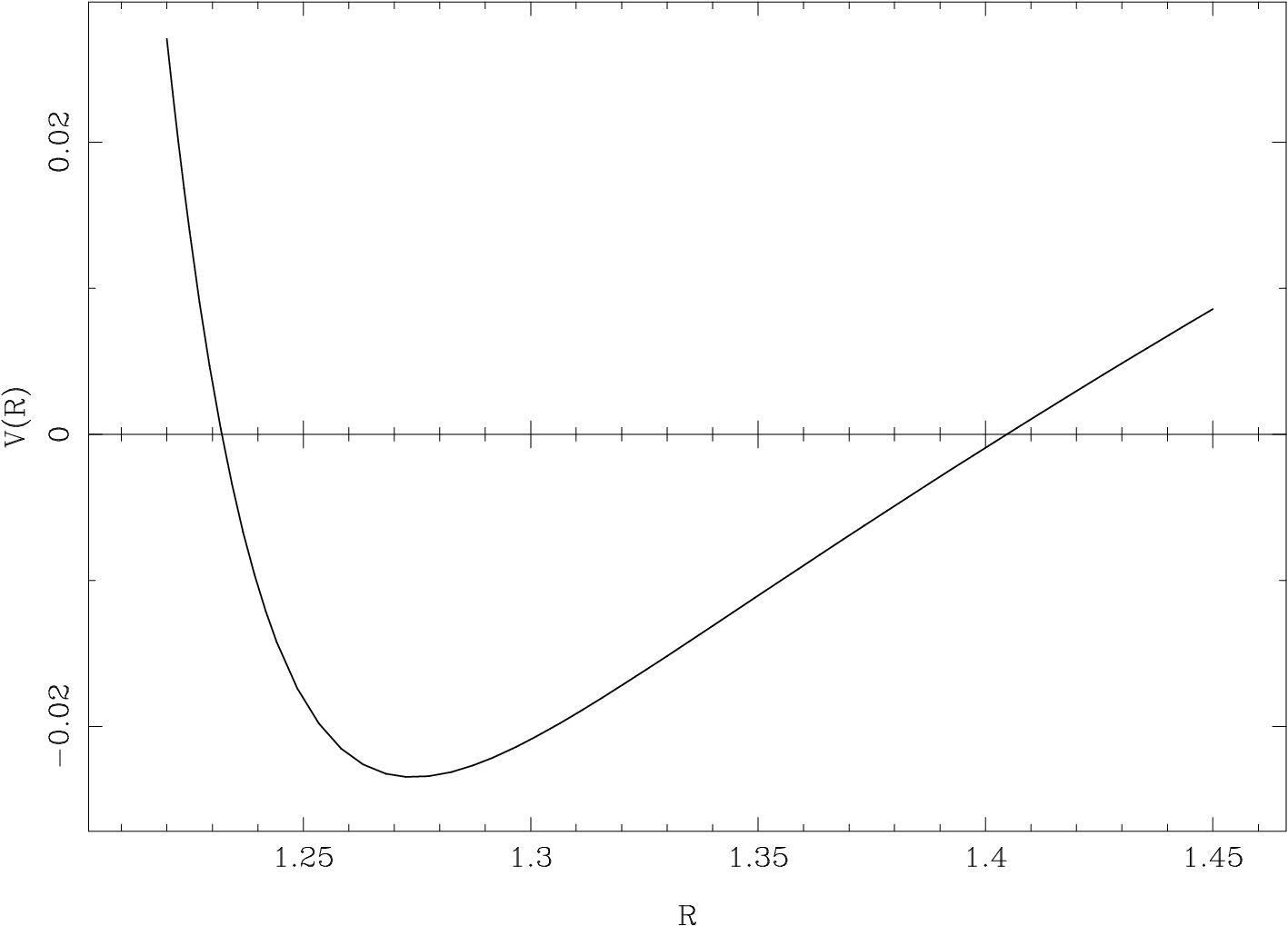,angle=-90,height=4cm,width=5cm}
\caption[]{\label{f4}
 Potential for the motion of the shell for the case of turning point 
radius of the order of one solar radius.}
\hspace{5cm}
\psfig{figure=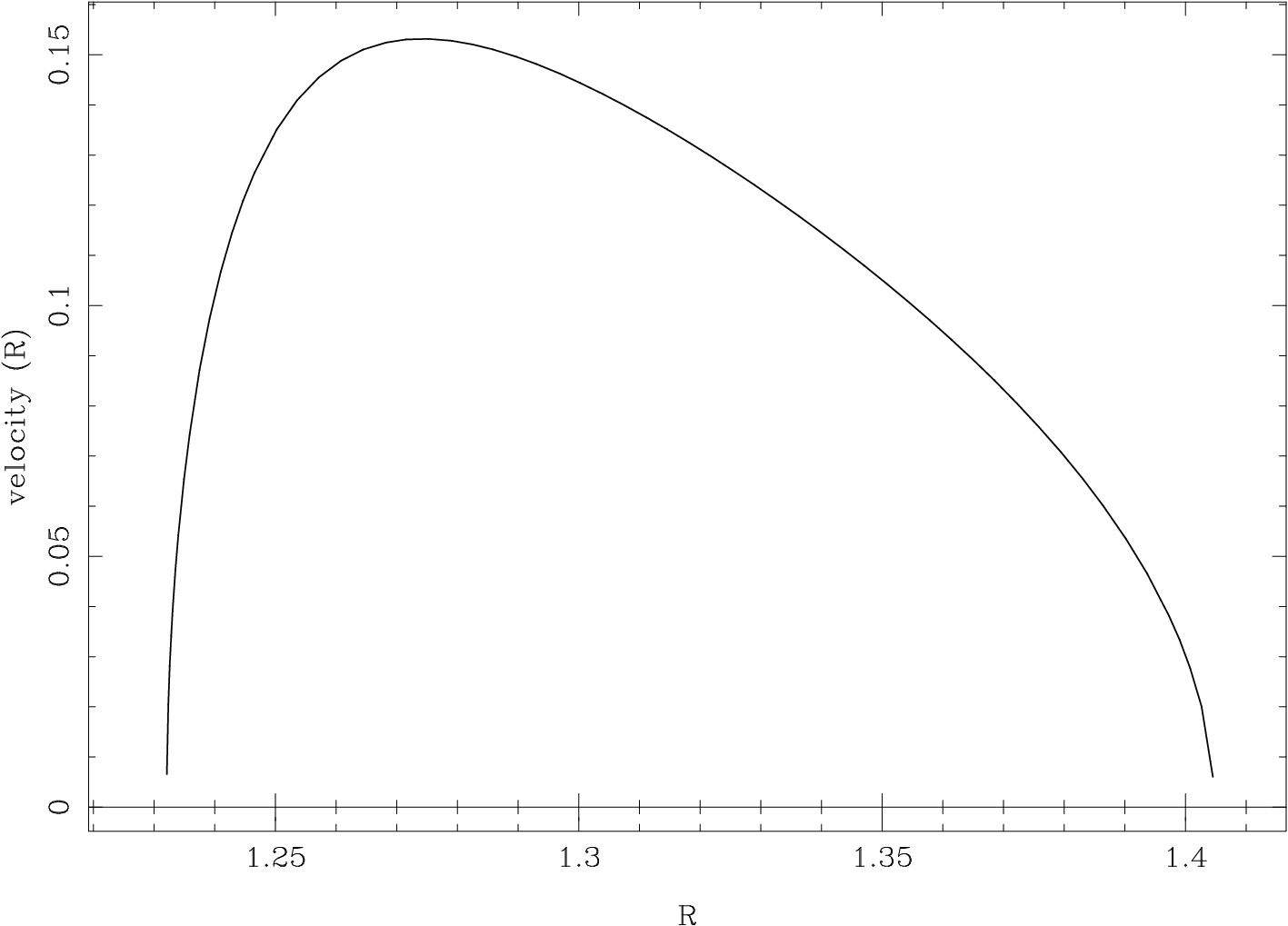,angle=-90,height=4cm,width=5cm}
\caption[]{\label{f5}
Shell velocity for the potential shown in Figure 4.}
\end{figure*}

 
Finally, for the surface density and the tangential pressure we obtain 
that their values change between the minimum radius and the 
maximum in the range 
\bea 
\sigma & \in  & [1.6429\,*\,10^{-13}, 1.231\,*\,10^{-13}]\,{\rm cm}^{-1} 
= \nonumber \\&& 
[2.21\,*\,10^{15}, 1.661\,*\,10^{15}]\, {{\rm g}\over{\rm cm^2}}, \\ 
p &  \in & [1.386\,*\,10^{-13}, 1.9823\,*\,10^{-17}]\,{\rm cm}^{-1} 
=  \nonumber \\ && 
[1.682\,*\,10^{36}, 2.406\,*\,10^{32}]\, {{\rm dyne}\over{\rm cm}}, 
\eea 
again the strong energy condition holds over the whole range and  
we obtain large
values for the surface density and the tangential stress.

\vfill 
\eject 

Static case. The equations of motion in our model also allow for a static 
solution. In this case we demand that the gravitational and stress forces 
cancel each other. This implies a constant gravitational mass of the shell, 
$M$, (see Eq.~(\ref{eq:mot1})). Notice that we do not need to specify
the equation of state for the matter on the shell. Our only claim is that
$M$ is constant so this model holds for all equations of state.

Since $\dot R =0$, then by the equation of motion (\ref{eq:mot2})
the potential must also be zero. Equating to zero the expression for the
potential, Eq.~(\ref{eq:pot1}), with the all masses taken as constants, we 
obtain an expression for the constant radius which satisfies this equation, 
that is, the static radius, $R_{\rm st}$: 
\be 
\tilde R_{\rm st}={{M_\odot\,\tilde M^2}\over{2\,R_\odot}}\,({{\tilde m_+
+\tilde m_-\pm 
\sqrt{\tilde M^2+4\tilde m_+\tilde m_-}}
\over{\tilde M^2-(\tilde m_+-\tilde m_-)^2}}). 
\ee 
Thus, at this radius the shell remains at rest. If we choose a shell of mass 
$\tilde M=0.1, M=1.98\,*\,10^{32}\,{\rm g}. $, 
surrounding a star with a mass such as that of the 
Sun, $\tilde m_-=1, m_-=1.98\,*\,10^{33}\,{\rm g}.$, and taking for 
the exterior space a gravitational mass $\tilde m_+=1.05, 
m_+=2.088\,*\,10^{33}\,
{\rm g}.$, then we obtain the 
value of $R_{\rm st}$ for which the shell is static surrounding this object,  
\be 
R_{\rm st}=5.7956\,*\,10^{-6} \,R_\odot=40.27 {\rm km}. 
\ee 
For the object in the interior we can choose any radius less that the static 
one and greater than Schwarzschild horizon of the space times or 
that of the shell, (these radii are, for the values of the gravitational 
masses chosen, $r_{\rm Sch_+}=3.09 {\rm km}.>r_{\rm Sch_-}
= 2.945 {\rm km}. > 
r_{\rm Sch_M}= 0.294 {\rm km}.$) It could be a neutron star 
with a typical radius of 
$R=15.12 {\rm km}.$ (S. Shapiro and P. Teukolsky, 1983).  

The stability analysis for this case has to be carried out to the second
order perturbations, as the first order one, Eq.~(\ref{eq:pert}), evaluated at
the static solution implies ${{\partial V}\over{\partial R}}|_{R_{st}}=0$ and
does not give us information about the perturbation. It seems more convenient
to analyze the perturbations starting from the second order differential 
equation for the motion:
\be
\ddot R = -{{\partial V}\over{2\,\partial R}},
\ee
which implies for the perturbation $\delta R$:
\be
\ddot{(\delta R)} = -{{\partial^2 V}\over{2\,\partial R^2}}|_{R_{st}}
\,\delta R,
\ee
an harmonic oscillation equation. It is important to notice that, even though
for the static case $\dot M=0$, this does not implies 
${{\partial M}\over{\partial R}}=0$. Actually, for each static radius, the
mass of the shell changes accordingly. In obtaining this variation of the
mass of the shell with respect to the static radius, we use again the static
condition, $V(R)=0$, taking it as an equation for 
$\tilde M=\tilde M(\tilde R_{st})$,
which results in a fourth order algebraic equation for $\tilde M$, with roots:
\be
\tilde M={R_\odot\over M_\odot}\,\sqrt{2\,\tilde R_{st}\,[\tilde R_{st}-
(\tilde m_+ + \tilde m_-)\,(M_\odot/R_\odot)
\pm\sqrt{\Delta}]},
\ee
where $\Delta=(\tilde R_{st}-2\,\tilde m_+\,(M_\odot/R_\odot))\,
(\tilde R_{st}-2\,\tilde m_-\,(M_\odot/R_\odot))$, and we are ignoring the 
clearly negative roots. From this last equation for $\tilde M$, we 
obtain that
\be
{{\partial M}\over{\partial R_{st}}}=\pm{{M\,(\tilde R_{st}\pm
\sqrt{\Delta})}\over{2\,R_{st}\,\sqrt{\Delta}}}. 
\ee
If we substitute this last expression in Eq.(\ref{eq:dpot}), that amounts
to evaluate ${{\partial V}\over{\partial R}}$ at the static radius, we
obtain that ${{\partial V}\over{\partial R}}|_{R_{st}}=0$, as expected. Now
we proceed to analyze the second derivative of the potential $V(R)$, which
is given by:
\bea
{{\partial^2 V}\over{\partial R^2}}=&
-{{2\,(\tilde m_+ + \tilde m_-)\,(M_\odot/R_\odot)}\over{{\tilde R}^3}}
-{{3\,{\tilde M}^2\,(M_\odot/R_\odot)^2}\over{2\,{\tilde R}^4}}
-({{6\,(\tilde m_+ - \tilde m_-)^2}\over{{\tilde M}^2}} +
{{{\tilde M}^2\,(M_\odot/R_\odot)^2}\over{2\,{\tilde R}^2}})\,
({{{\tilde M}^\prime}\over{\tilde M}})^2 +&\cr
&
{{\tilde M\,{\tilde M}^\prime(M_\odot/R_\odot)^2}\over{{\tilde R}^3}}
+ 2\,({{(\tilde m_+ - \tilde m_-)^2}\over{{\tilde M}^2}} -
{{{\tilde M}^2\,(M_\odot/R_\odot)^2}\over{4\,{\tilde R}^2}})\,
({{{\tilde M}^{\prime\,\prime}}\over{{\tilde M}^2}}),&
\eea 
where ${}^\prime$ stands for the derivative with respect to $\tilde R$. 
Computing the second derivative of $M$ and substituting in this last 
expression that, as we mentioned above, is equivalent to evaluating the
potential at the static solution (the results have been checked 
using MapleV-4, which has been used also in obtaining the figures), 
we obtain that
\be
{{\partial^2 V}\over{\partial R^2}}|_{R_{st}}=0.
\ee
So, the perturbation equation for this case reduces to
\be
\ddot{(\delta R)}=0,
\ee
which in turn implies that the radial perturbations for the static shell
are given by
\be
\delta R = a\, \tau +b,
\ee
with $a$ and $b$ arbitrary constants, so they grow linearly and we thus 
conclude that the static shell is unstable under radial perturbations. 

These results allow us to conclude that the potential for the static case
has no minimum around the static radius but an inflection point so it 
has the form of a large plateau
around that static radius and that under radial perturbations, which will
grow linearly with respect to the proper time of the shell, will tend either 
to collapse or expand, depending on the sign of the initial conditions imposed
for the perturbation. If $\dot{\delta R}(\tau_0)=a>0$, it expands, and if
$\dot{\delta R}(\tau_0)=a<0$, it collapses.

\vspace{7mm} 
\section{Conclusions} 
 
In the present work we have presented the analysis of the equations of motion 
for a shell of perfect fluid using the formalism of General Relativity. We 
have chosen a model where the effects of the combined action of two type of 
forces acting on the shell are shown, namely the gravitational attraction 
versus a force due to the tangential stresses. Even 
though the problems has been
reduced to a set of two coupled partial differential equations, an 
analytic solution was not found. Nevertheless the equations are very 
tractable and from the properties of the potential energy we have been able to
present several types of motion allowed in our model, which depends on five 
parameters. We presented 
cases for oscillatory motion for masses of the order of 
a Solar mass and small radius, and for a radius of the order of Solar radius, 
which implies large masses. With an appropriate selection of the parameters 
for any value of mass or radius, such a motion can be found. We 
have also presented other two cases allowed by our model, one which could be 
interpreted as explosion, an ejection of a shell starting from rest due to the
force associate with the tangential pressure, and another where that force 
equals the gravitational one, allowing the shell surrounding a cosmic body 
to be static. The matter in the shell is associated with an 
unusual equation of state, 
which does not violate any energy 
condition, so we can say that it is not an ``exotic" type of matter but still
unlikely to describe a realistic one, so it is actually a new, 
mathematically correct solution but the reader should be aware of the type of
matter used in this model.

This last remark though, does not apply to the last case presented, the
static one, where we have obtained the model without any reference to an
equation of state, so it holds for all of them. We also showed that
the static shell is an unstable configuration under radial perturbations which
grow linearly with respect to the proper time of the shell. 
 
We can incorporate more parameters to describe other type of objects. For 
instance, we can take into account the radiation emitted 
from the inner body to the shell and the interaction of the shell with the 
medium, the radiation pressure and the friction type force respectively, 
considered in the classical formalism. This case could be obtained working 
with the matching conditions on 
the shell of a Vaidya universe inside (Vaidya P. C. 1951), and a Friedmann - 
Robertson - Walker dust universe outside (see Weinberg S. 1972). Also some 
deviations from sphericity could be considered, so that the matching 
conditions could be analyzed for an axisymmetric type of space times. This 
case is expected to be highly unstable, so the point would be to study the
final state of that model. Finally, it would be of 
interest to study the motion
of shells composed of different types of matter, such as one 
constructed from a scalar
field. These ideas are currently under investigation. 

We think that the model presented elucidates several features of the possible
types of motion of the shell.  
 
I want to acknowledge the project IN105496 from DGAPA-UNAM for partial support
during the development of the present work. Also, I want to thank H. Quevedo 
as well as M. Salgado, D. Sudarsky and T. Zannias for fruitful 
comments and discussions about this work. Finally, I thank the anonymous 
referee for helpful comments and suggestions.

\vspace{7mm} 
\section{Appendix} 
 
In this appendix we present a review of the description on the deduction of 
the equations of motion of the shell, Eqs.(\ref{eq:mot1}, \ref{eq:mot2}). The 
four - dimensional space-time is taken to be composed of two parts 
${\cal M}_-$ and ${\cal M}_+$, separated by a boundary $\Sigma$. The main 
goal in this analysis consists in showing how the geometry on $\Sigma$ is 
determined by that of the space-time in which it is contained. Then,  
demanding continuity of the line elements on $\Sigma$, and analyzing the jump 
on the curvature due to the presence of the shell, to obtain the ``matching" 
conditions or, as seeing in the present work, the equations of motion of the 
shell.  
 
 The first condition for joining the two regions is that the line element 
of the region ${\cal M}_-$ should equal the line element of ${\cal M}_+$ at 
$\Sigma$, that is, both must be equal to ${ds^2}_\Sigma$. 
 
The metric on the shell is described by $\gamma_{ab}$, (Latin indexes are 
0,2,3, Greek  0,1,2,3). It is related to the metric of the space-time by 
$e^\alpha_{(a)}\,e^\beta_{(b)}\,\gamma^{ab}=
g^{\alpha\beta}+ n^\alpha\,n^\beta$, where $n^\alpha$ is a unit 
4-dimensional 
vector normal to $\Sigma$, and the $e^\alpha_{(a)}$ are the projectors on the 
hypersurface: $e^\alpha_{(a)}A^a=A^\alpha, e^\alpha_{(a)}A_\alpha = A_a$, for 
an arbitrary vector {\bf A}. The 
4-dimensional velocity is 
$u^\alpha\equiv {{dx^\alpha}\over{d\tau}}$ and the 
4-dimensional acceleration 
${{\delta \, u^\alpha}\over{\delta \tau}}={u^\alpha}_{;\mu}u^\mu$, (semicolon 
denotes covariant derivative with respect to $g_{\alpha\beta}$).

One of the most important geometric quantities in the embedding is the 
extrinsic curvature defined by $K_{ab}=n_{\beta}\,{e^\beta}_{(a);\alpha}\,
{e^\alpha}_{(b)}$. Recalling that ${e^\beta}_{(a)}\,n_\beta=0$, it is not hard
to see that the extrinsic curvature is related to the acceleration by  
$$ 
K_{ab}u^a\,u^b= n_\alpha\,{{\delta \, u^\alpha}\over{\delta \tau}}.
\eqno{(A.1)} \label{eq:kuu} 
$$  
So the extrinsic curvature 
projected on the 3-dimensional velocity equals to the projection of the 
4-dimensional 
acceleration on the normal. The first matching condition refers to the 
difference betwwen the extrinsic curvature projected 
from ${\cal M}_+$ and the 
one obtained projecting from ${\cal M}_-$. Let us denote this difference by 
$\mu_{ab}$: 
$$ 
\mu_{ab}=\left[ K_{ab} \right]=K_{ab}|^+ - K_{ab}|^-. 
\eqno{(A.2)} \label{eq:muab} 
$$ 
where a quantity in square brackets denotes that difference of the 
projections: $\left[ A \right]=A^+ - A^-$. From Eq.~(A.2) together 
with Eq.~(A.1), we can obtain an expression for the difference of 
the acceleration projected on the normal from the two regions of the 
space-time: 
$$ 
\left[ n_\alpha\,{{\delta \, u^\alpha}\over{\delta \tau}} \right]=
\mu_{ab}u^a\,u^b,\eqno{(A.3)} \label{eq:nac} 
$$ 
 
Another important geometric quantity is the symmetric 3-tensor $S_{ab}$, 
defined by the ``Lanczos equations"   
$$ 
\mu_{ab} - \gamma_{ab}\,\mu = 8\,\pi\,S_{ab}, \nonumber 
$$ 
or equivalently, $\mu_{ab}=8\,\pi\,(S_{ab} - {1\over2}\gamma_{ab}\,S)$, 
($\mu=\gamma^{ab}\mu_{ab}, S=\gamma^{ab}S_{ab}$). 
The quantity $S_{ab}$ can be considered 
to be the {\it surface energy tensor} of the shell, as it is the limit of the 
integral of the stress energy tensor through the thickness of the shell when 
this thickness tends to zero (Israel 1966). We will take the matter on the 
shell to be described by a perfect fluid,  
$$ 
S_{ab}=(\sigma + p)\,u_a\,u_b + p\gamma_{ab}, \eqno{(A.4)} 
$$ 
where $\sigma$ is the matter-energy density on the shell, $p$ is the 
tangential pressure of that matter, and $u^a$ stands for the 
3-dimensional time-like 
velocity vector, $\gamma_{ab}u^a\,u^b=-1$. Then, the equation Eq.~(A.3) for 
the difference of the projections of the acceleration becomes 
$$ 
\left[ n_\alpha\,{{\delta \, u^\alpha}\over{\delta \tau}} \right]= 
{M\over{R^2}}  + 8 \pi p, \eqno{(A.5)}, 
$$ 
where $M$ (not necessarily constant) is the gravitational mass of the shell, 
defined by $M=A \sigma$, with $A$ the area of the shell, $A=4\pi R^2$. 
Eq.~(A.5) is the first of the matching equations, which tells us that a 
discontinuity of the normal component of the acceleration is related to the 
matter present on the shell.  
 
There are two other matching equations which can be deduced starting from the
relations of the components of the Riemann tensor in the space-time with the 
one on $\Sigma$. These relations are called the Codazzi-Mainardi equations: 
$$ 
{}^4{R^n}_{bcd}=K_{bc|d} - K_{bd|c},\nonumber 
$$ 
and the Gauss-Codazzi equations: 
$$ 
{}^4R_{abcd}={}^3R_{abcd} + K_{ac}\,K_{bd} - K_{ad}\,K_{bc},\nonumber 
$$ 
where $|$ denotes covariant derivative on $\Sigma$, {\it i. e.}, 
$\gamma_{ab|c}=0$. Multiplying the Codazzi-Mainardi equations by 
$\gamma^{bc}$,
and the Gauss-Codazzi by $\gamma^{ac}\,\gamma^{bd}$, we get a set of four 
equations, valid on each side of the shell: 
$$ 
G_{\mu\nu}\,n^\mu\,{e^\nu}_{(a)}\,|^{\pm}=({K^b}_{a|b} - K,_{a})|^{\pm}, 
\nonumber 
$$ 
$$ 
2G_{\mu\nu}\,n^\mu\,n^\nu\,|^{\pm}={}^3R + (K^2 - K_{ab}\,K^{ab})|^{\pm},
\nonumber 
$$ 
 
With these definitions and the Einstein equations, 
$G_{\mu\nu}=8\,\pi\,T_{\mu\nu}$, the difference of the projections of the 
stress-energy tensor evaluated on the shell are: 
$$ 
\left[ T_{\mu\nu}\,n^\mu\,u^\nu\right]  = u^{(a)}\,{S^b}_{a|b}, \eqno{(A.6)} 
  \label{eq.II} 
$$ 
$$ 
2\left[ T_{\mu\nu}\,n^\mu\,n^\nu \right]  =  -S^{ab}(K^+_{ab}+K^-_{ab}), 
\eqno{(A.7)}   \label{eq.I} 
$$ 
which are the other two matching equations, relating the projections of the 
stress energy tensor on the normal and on the velocity and 
on the normal on both 
indexes to the matter distribution on the shell. Eq.~(A.6) is the equation of 
energy balance that tells us how the fluxes of matter-energy from both sides 
determine the dynamics of the shell. On the other hand, Eq.~(A.7) has no 
clear physical interpretation (see below). 
 
For the matter-energy of the shell described by a perfect fluid, and taking 
into account the fact 
that $\dot{\sigma}\equiv{{d\sigma}\over{d\tau}}$ and that 
${u^b}_{|b}={1\over{\sqrt{-\gamma}}}((\sqrt{-\gamma}\,u^b)_{,b}$, with 
$\gamma\equiv det(\gamma_{ab})$, as well as that for spherical symmetry, the 
line element of the shell is given by Eq.~(\ref{eq:lel}). The shell $\Sigma$ 
is a 3 dimensional manifold, characterized by a line element 
$$ 
{ds^2}_\Sigma=-d\tau^2+R^2(\tau)d\Omega^2, \eqno(A.8) 
$$ 
with $R(\tau)$ the radius of the shell. 
Thus we can obtain that ${u^b}_{|b}={{\dot A}\over A}$, with $A$ the area of 
the shell defined above, the energy balance equation, Eq.~(A.6), takes the 
form 
$$ 
\dot{M}+p\dot{A}=4\,\pi\,R^2\,\left[T_{\mu\nu}\,n^\mu\,u^\nu\right], 
\eqno(A.9) 
$$ 
which is the form of the second matching condition 
(N{\'u}{\~n}ez and Oliveira 1996). 
The interpretation as the equation of motion 
for the shell looks particularly clear
in this form. Here it is important to stress the fact that 
with spherical symmetry and a perfect fluid, the first matching condition 
(A.5) can be integrated without further assumptions, using Eq.~(A.9), 
to yield a quadratic first order equation for $R(\tau)$, Eq.~(\ref{eq:mot2}). 
For details on this deduction see (N{\'u}{\~n}ez and Oliveira 1996). 
 
Finally, the third matching condition (A.7) which, using Eq.~(A.1) and 
$S^{ab}$ as a perfect fluid, Eq.~(A.4), tells us that the sum of the 
projections of the acceleration on the normal, from each side of $\Sigma$ is 
given by 
$$ 
\sigma\,\{ n_\alpha\,{{\delta \, u^\alpha}\over{\delta \tau}} \}\equiv 
\sigma\,(n_\alpha\,{{\delta \, u^\alpha}\over{\delta \tau}}|^+ + 
n_\alpha\,{{\delta \, u^\alpha}\over{\delta \tau}}|^- ) = 
-2\left[ T_{\mu\nu}\,n^\mu\,n^\nu \right]-p\,\{K_{ab}u^au^b + K\,\}. 
\eqno(A.10) 
$$ 
As mentioned above, this equation has no clear physical meaning. As a 
matter of fact, it can be shown that, at least for the spherical case with the
matter-energy of the shell described by a perfect fluid, Eq.~(A.4), this last 
equation Eq.~(A.11) is the same as the matching equation Eq.~(A.5), so we do 
not have to consider it in the analysis of the motion of the shell. For 
details on the demonstration of this equality see (N{\'u}{\~n}ez, 1996). 

In this way, the motion equations for the shell reduce to a set of two first 
order equations, namely Eqs.~(\ref{eq:mot1}, \ref{eq:mot2}). The equation for 
the change in the advanced time can be obtained from the equation for the 
change in the radius, using the normalization equation for the cuadrivelocity.
The problem is completed with a equation of state for the 
matter-energy of the 
shell, $p=p(\sigma)$.
\vfill
\eject 
\centerline{REFERENCES} 

\vskip1pc 
\noindent 
De la Cruz, V., \& Israel, W. 1967, Il Nuovo Cimento, LI A, 3, 744 
 
\vskip1pc 
\noindent 
Israel, W. 1966, Il Nuovo Cimento, 44 B, 1 
 
\vskip1pc 
\noindent 
 ------. 1967, 48 B, 463 
 
\vskip1pc 
\noindent 
Lichnerowicz A. 1955, Th{\'e}ories Relativistes de la Gravitation et de l´Electromagnetisme,  Ed. Masson et Cie, Paris. 
 
\vskip1pc 
\noindent 
N{\'u}{\~n}ez, D. 1996, in preparation 
 
\vskip1pc 
\noindent 
N{\'u}{\~n}ez, D., and de Oliveira, H. P. 1996, Phys. Lett. A, 214, 227 
 
\vskip1pc 
\noindent 
N{\'u}{\~n}ez, D., de Oliveira, H. P and Salim J. 1993, Class. Quantum Grav. 10, 1117 
 
\vskip1pc 
\noindent 
Ostriker, J. P., \& Gunn, J. E. 1971, ApJ, 164, L95 
 
\vskip1pc 
\noindent 
Shapiro S., and S. Teukolsky S. 1983, ``Black holes, white dwarfs and neutron stars", Wiley - Inter science publications, New York 
 
\vskip1pc 
\noindent 
Vaidya P. C. 1951, Proc. Indian Acad. Sci. A 33, 264 
 
\vskip1pc 
\noindent 
Weinberg S. 1972, ``Gravitation and Cosmology. Principles and applications of the General Theory of Relativity", Wiley \& Sons, New York 
 
\end{document}